\input epsf
\input harvmac
\vglue 2cm
\centerline{\bf{Geometric Physics}}
\vglue 2cm
\centerline{ Talk presented at the ICM `98}
\vglue 1cm
\centerline{Cumrun Vafa}
\centerline{ Lyman Laboratory of Physics}
\centerline{Harvard University}
\centerline{Cambridge, MA 02138, USA}
\vglue 1cm
\centerline{\bf ABSTRACT}
\vglue 1cm
Over the past two decades there has been growing interaction
between theoretical physics and pure mathematics.  Many of these
connections have led to profound improvement in our understanding
of physics as well as of mathematics.  The aim of my talk is to
give a non-technical review of some
of these developments connected with string theory.

The central phenomenon in many of these links involves
the notion of {\it duality}, which in some sense
 is a non-linear
infinite dimensional generalization of
the Fourier transform.
  It suggests that two physical systems
with completely different looking properties are nevertheless
isomorphic if one takes into account ``quantum geometry'' on both
sides.  For many questions one side is simple (quantum
geometry is isomorphic to classical one) and the other
is hard (quantum geometry deforms the classical one).  The
equivalence of the systems gives rise to a rich
set of mathematical identities.
One of the best known examples of duality is
known as ``mirror symmetry'' which relates
topologically distinct pairs of Calabi-Yau manifolds and has applications in
enumerative geometry.  Other examples involve highly
non-trivial ``S-dualities'' which among other things have
found application to the study of smooth four manifold invariants.
There have also been applications to questions of quantum gravity.
In particular certain
properties (the area of the horizon)
 of black hole solutions to Einstein equations
 have been related to growth of the
cohomology of the moduli space of certain minimal submanifolds in a Calabi-Yau
threefold.

A central theme in applications of dualities
is a physical interpretation of singularities of manifolds.
The most well known example is the $A-D-E$ singularities of the
$K3$ manifold which lead to $A-D-E$ gauge symmetry in the physical
setup.  The geometry of contracting cycles is a key ingredient
in the physical interpretation of singularities.
More generally, singularities of manifolds encode
universality classes of quantum field theories.
This leads not only to a deeper understanding of the
singularities of manifolds but can also be used
to ``geometrically engineer'' new quantum field theories
for physics.
%Through this method many non-trivial properties
%of quantum field theories get related to classical symmetries
%of geometry.  An example of this application
%(in the context of Montonen-Olive duality) is
%a simple geometric interpretation of the
% modular properties of the generating function of the Euler
%class of moduli space of instantons on four manifolds.
%\Date{8/98}
%\draft
\newsec{Introduction}
The history of physics and mathematics is greatly interconnected.
Sometimes new mathematics gets developed in connection with understanding
physical questions (for example the development of Calculus was not independent
of the questions raised by classical mechanics).  Sometimes new physics
gets developed from known mathematics (for example general theory of relativity
found its natural setting in the context of Riemannian geometry).  I believe we
are now witnessing
perhaps an unprecedented depth in this interaction between the two disciplines.
It is thus a great pleasure to explain some of the recent progress which
has been made in our understanding of quantum field theories, string
theory and quantum gravity to a mathematical audience.  The works I will be
explaining here is a result of the work of many physicists
and mathematicians.\foot{I will not make any attempts to present a complete
list of references to all the relevant literature, though some illustrative
references, in the spirit of the presentation here will be given.}

Many of the key elements in these recent advances have a deep mathematical
content.  These involve new predictions for answers to some very difficult
mathematical questions as well as new interpretations of some old mathematical
results.  It also sometimes hints at the existence
of whole new branches of mathematics which does not exist yet.

In preparing this talk, I have had to make some choices.  First of all
I have had to decide which topics to cover and which ones to leave out.
This has been very difficult because there are many interesting interaction
points between theoretical physics and pure mathematics today, and
unfortunately
I only have a very limited time here.   My choice was motivated by the degree
of my
familiarity with the subject as well as
by attempts at trying to give a unified exposition of the
seemingly unrelated topics.
  Secondly I have had to assume
a certain level of familiarity of this mathematical audience with physics.
This is also unavoidable, if we are to make any connection to interesting
new developments.  However, I have tried to make this assumption in the weakest
possible sense.  Thirdly I have chosen a list of questions
which I find interesting for physics which I hope the mathematicians
will help us solve.

The organization of my talk is as follows:  In section 2 I will
describe the basic notion of {\it duality} which is the key notion
in recent advances.  In sections 3-5 I give examples of
dualities.  Section 3 is devoted to a review of what mirror symmetry
is.  Section 4 explains the physical interpretation of singularities
of certain manifolds.  Section 5 is devoted to the notion
of black hole entropy and what duality predicts about that.  Section 6
is devoted to a list of questions which I raise in connection
with the topics discussed.

\newsec{What is meant by Duality?}
I will try to define a very general notion of duality first, a priori
nothing to do with physics, and then
try to be a little more particular in what it means in the physical context.

Suppose we have two classes of objects.  Moreover suppose these
two objects satisfy identical properties.  Then in a mathematical
context they usually will be called {\it isomorphic}.  Very often
this is a trivial isomorphism.  For example if a property of geometry
on a 2 dimensional plane is true, it will also be true for the mirror
reflection of the same geometry (Fig. 1).

\bigskip
%\midinsert
\epsfxsize 2.truein
\epsfysize 2.truein
\centerline{\epsfxsize 2.truein \epsfysize 2.truein\epsfbox{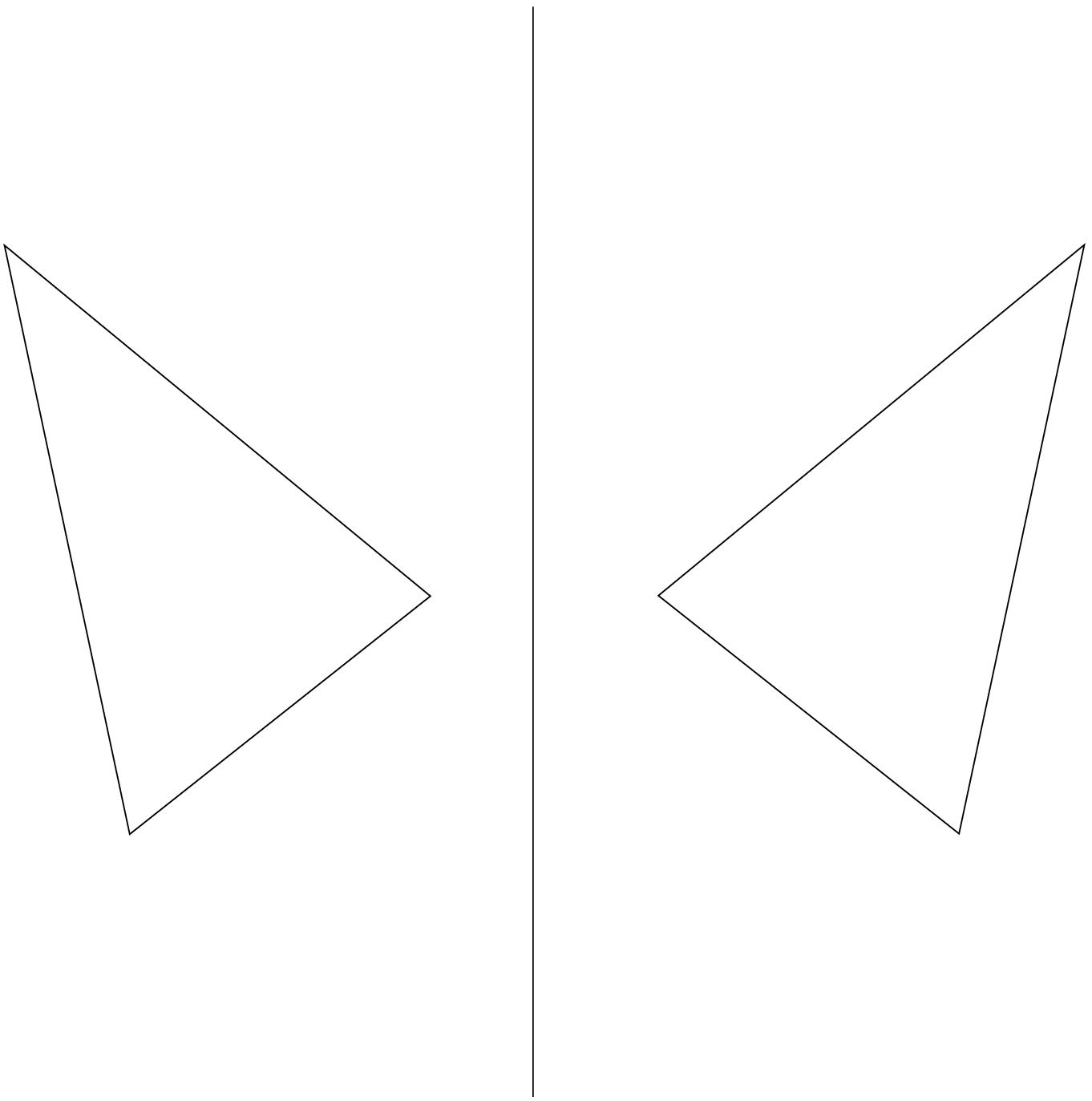}}
\leftskip 2pc
\rightskip 2pc
\noindent{\ninepoint\sl \baselineskip=8pt {\bf Fig.1}: {\rm Reflection
on the plane is an example of a ``trivial'' duality.}}

However there are times where the fact that
the objects and operations are isomorphic is less trivial, because
the maps between these two classes of objects is not so trivial.
As an example, suppose we wish to solve a linear differential equation
of the form
$$F=\sum_k a_k{d^k\over dx^k}\psi(x)=0$$
with constant coefficients $a_k$.  Consider instead the {\it polynomial}
equation in one variable $p$:
$$G=\sum a_k (ip)^k=0$$
Apriori the two problems seem unrelated.  In fact the
second problem on the face of it sounds much simpler.  However, as is well
known the two problems are related by Fourier transform, and the general
solution to the first problem is given by
$$\psi(x)=\int dp \phi(p) {\rm exp}(ip x)\delta(G(p))$$
This isomorphism of functions in $x$ and functions in $p$
with the map between them being Fourier transform allows us to solve
a `hard' problem in the $x$ space setup in terms of an easy problem
in the $p$ space setup.  Isomorphisms of this type which are
{\it non-trivial} we will call {\it dualities}.  As it is clear
from this example dualities will be very useful in solving problems.
Dualities very often transform a difficult problem in one setup
to an easy problem in the other.  In some sense very often the very
act of `solving'
a non-trivial problem is finding the right `dual' viewpoint.

Now I come to specializing this idea in the context of a physical
system.  Consider a physical system $Q$
(which I will not attempt to define).  And suppose this system depends
on a number of parameters $[\lambda_i]$.   Collectively we denote
the space of the parameters $\lambda_i$ by ${\cal M}$
which is usually called the moduli space of the coupling constants of
the theory.
The parameters $\lambda_i$ could for example define
the geometry of the space the particles propagate in, the charges
and masses of particles, etc.  Among these parameters there is a parameter
$\lambda_0$ which controls how close the system is to being a classical
system (the analog of what we call $\hbar$ in quantum mechanics).  For
$\lambda_0$ near zero
we have a classical system and for $\lambda_0\geq 1$ quantum effects typically
dominate the description of the physical system.  Typically physical
systems have many observables which we could measure. Let us denote
the observables by $\cal O_\alpha$.  Then we would be interested in
their correlation functions which we denote by\foot{One could attempt
to define a physical system by an infinite dimensional bundle over ${\cal M}$
where the fiber space is identified with the space of observables
${\cal O_\alpha}$, together with a rank $n$ multi-linear map from the
fiber to ${\bf C}$, for each $n$, satisfying some compatibility conditions.}
$$\langle {\cal O}_{\alpha_1}...{\cal O}_{\alpha_n}\rangle =
f_{\alpha_1 ...\alpha_n}(\lambda_i)$$
Note that the correlation functions will depend on the parameters
defining $Q$.
The totality of such observables and their correlation functions
determine a physical system.  Two physical systems $Q[{\cal M},{\cal
O}_\alpha]$,${\tilde Q}[\tilde {\cal M},\tilde {\cal O}_\alpha]$
are dual to one another
if there is an isomorphism between ${\cal M}$ and $\tilde{\cal M}$
and ${\cal O}\leftrightarrow {\tilde {\cal O}}$
respecting all the correlation functions.   Sometimes this isomorphism
is trivial and in some cases it is not.  We are interested in the cases where
this isomorphism is non-trivial.  In such cases typically what happens
is that a parameter which controls quantum corrections $\lambda_0$ on one
side gets transformed
 to a parameter
${\tilde \lambda_k}$ with $k\not= 0$ describing some classical aspects of the
dual
side.
 This in particular implies
that quantum corrections on one side has the interpretation on the dual
side as to how correlations vary with some classical concept such as geometry.
This allows one to solve difficult questions involved in quantum
corrections in one theory
in terms of simple geometrical concepts on the dual theory.
This is the power of duality in the physical setup.  Mathematics parallels
the physics in that it
turns out that the mathematical questions involved in computing quantum
corrections
in certain cases is also very difficult and the questions involved
on the dual side are mathematically simple.  Thus non-trivial
duality statements often lead to methods of solving certain difficult
mathematical problems.

One should note, however, that very rarely can one actually {\it prove}
(even in the physics sense of this word) that two
given physical systems are dual to one another. Often the existence of
dualities between two systems is guessed at
based on some physical consistency arguments.  Testing
many non-trivial consequences of duality conjectures leads us to believe
in their validity.
In fact we have
observed that duality occurs very generically, for reasons we do not fully
understand.   This lack of deep understanding of duality is not unrelated to
the fact that it leads to solutions of otherwise very difficult problems.
At the mathematical level, evidence for duality conjectures amounts
to checking validity of proposed solutions to certain difficult
mathematical problems.

In the next three sections I will consider examples of duality and
some of its mathematical consequences.
In section 3 we will start with the best
understood duality known as mirror symmetry, which relates
string theory on one target manifold with another.  In section 4
we discuss
how singularities of the geometry get related to gauge bundles for the dual
theory.  In section 5 we discuss a dual description of black hole geometry
which is intimately related to properties of minimal submanifolds in
Calabi-Yau manifolds.

\newsec{Mirror Symmetry}
String theory, which is the only known consistent framework for a
quantum theory of gravity, involves the study of quantum properties of one
dimensional
extended objects.  The spacetime picture corresponds to a two dimensional
Riemann surface $\Sigma$  mapped to a target spacetime Riemannian manifold $M$.
The sliced Riemann surfaces give the picture of strings propagating in time
(Fig. 2).

\bigskip
%\midinsert
\epsfxsize 2.truein
\epsfysize 2.truein
\centerline{\epsfxsize 4.truein \epsfysize 2.truein\epsfbox{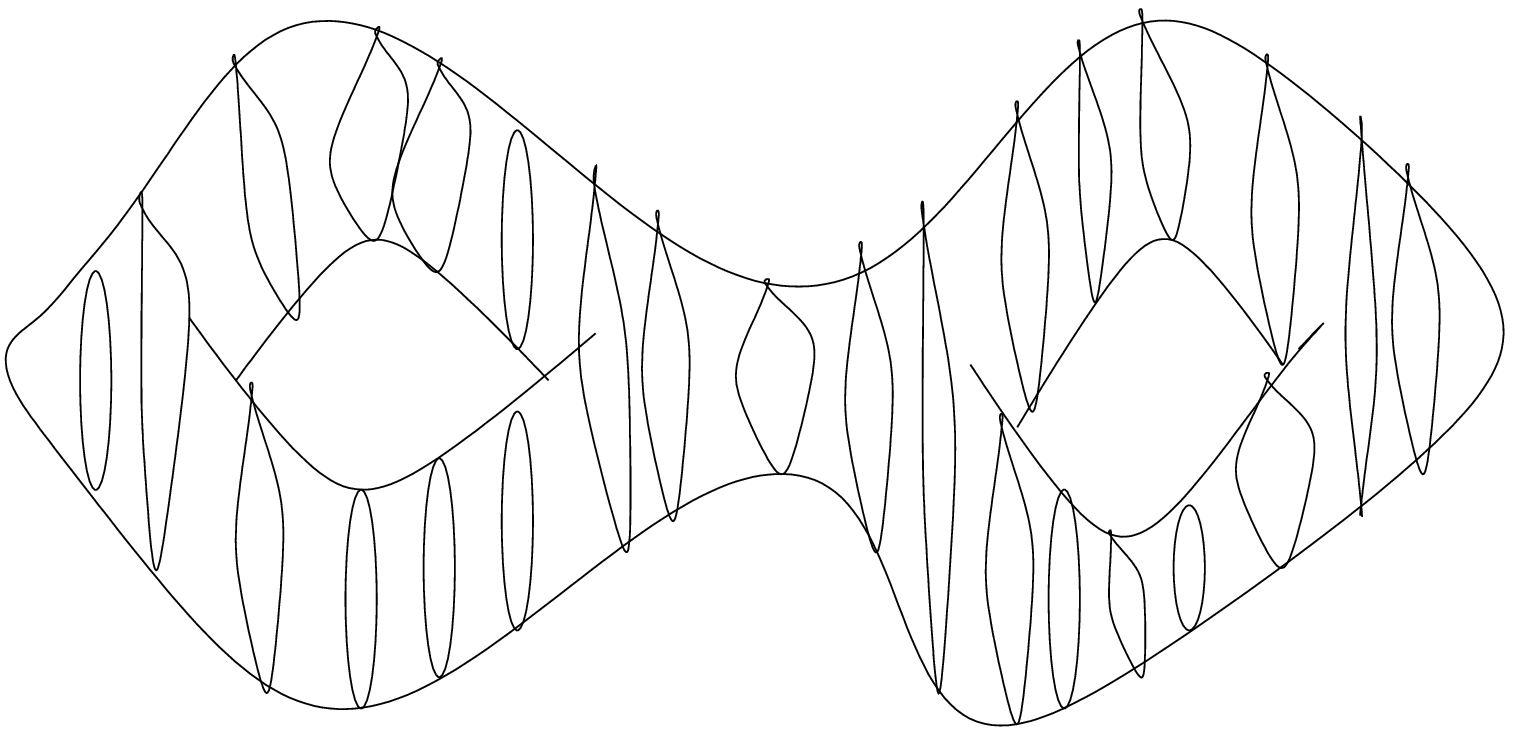}}
\leftskip 2pc
\rightskip 2pc
\noindent{\ninepoint\sl \baselineskip=8pt {\bf Fig.2}: {\rm
Strings propagating in spacetime span a Riemann surface known as the
worldsheet.}}

In string theory we are instructed to ``sum'' over all such maps
$$\phi: \qquad \Sigma \rightarrow M$$
weighted with $\rm exp (-S(\phi))$ where $S(\phi)$ denotes the
integral
$$S(\phi)= \int_\Sigma |d\phi|^2$$
where we use the metric on $M$ to define $|d\phi|^2$.  (For superstrings
which is the case of most interest, there are also some fermionic
fields, which I suppress in this discussion.)

One of the most amazing properties of string theory is that
strings moving on one manifold may behave identically with
strings moving on a different manifold.  Any pair of manifolds $M_1$
and $M_2$
which behave in this way are called mirror pairs. Of course this
would be a trivial duality if $M_1$ and $M_2$ are isomorphic
Riemannian manifolds.  The interesting dualities arise when
$M_1$ and $M_2$ are distinct Riemannian manifolds.  In some
cases $M_1$ and $M_2$ are topologically the same, but in some
cases they are distinct even topologically.  In such cases the
equivalence of the two manifolds for
string theory will be only a statement about
correlation functions after summing over {\it all} maps $\phi$.  The act
of summing over all maps $\phi$ is what we mean by the quantum theory.
So only in the quantum theory, i.e. after summing over
all $\phi$ the two computations would be related (i.e. we should not
try to compare individual maps).  The parameter controlling
the significance of quantum corrections, for a fixed genus surface
$\Sigma$,
is the volume of $M$, $V(M)$.  In particular, the parameter we called
$\lambda_0$ in
the previous discussion in this case is $\lambda_0=1/V(M)$ (and thus
in the large volume limit the quantum corrections are suppressed).

The simplest example of mirror symmetry corresponds to choosing
$M_1$ to be a circle of circumference $L$ and $M_2$ to be a circle
of circumference $1/L$.  This is a case of mirror symmetry
which can be rigorously proven (see \ref\vab{C. Vafa, in
{\it Geometry, Topology, and Physics for Raoul Bott}, ed. S.-T. Yau,
International Press 1995.} for a review).  However here we will just illustrate
why such a statement is not unreasonable.

This statement would definitely be unreasonable for point particle
theories:  If we consider a particle in a circle of size $L$, the momentum
states are quantized as the allowed wave functions
$$\psi_n(x)={\rm exp} (2\pi inx/L)$$
compatible with the invariance under $x\rightarrow x+L$ gives the
spectrum of allowed momenta
(which for massless particles is the same as energy)
to be $n/L$, where $n\in {\bf Z}$.  If we consider the circle of circumference
$1/L$
the allowed energies are now $nL$.  Thus the energy spectrum of the two
theories do not match.  The story changes dramatically for strings:
We will still have the same excitations as in the point particle case,
after all the string mapped to a point looks like a point particle.
However we have in addition other states corresponding to winding
states of the string around the circle.  Consider the first circle
of circumference $L$ and assume a string wraps around it $m$ times, then
its energy is $mL$ (I am working in units where the string tension is one).
Now the full spectrum of momentum and winding states does have $L\rightarrow
1/L$ symmetry where in the process momentum states get exchanged with
winding states (Fig. 3).

\bigskip
%\midinsert
\epsfxsize 2.truein
\epsfysize 2.truein
\centerline{\epsfxsize 2.truein \epsfysize 2.truein\epsfbox{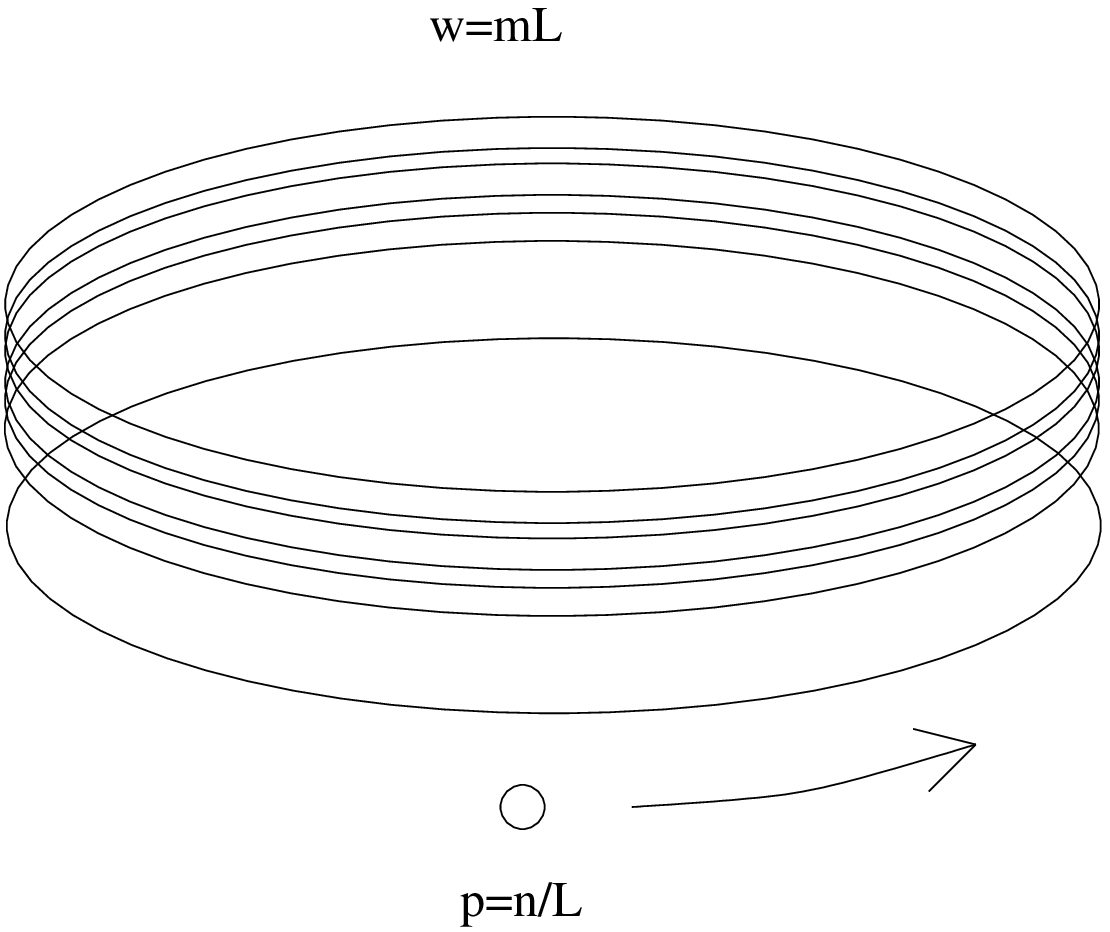}}
\leftskip 2pc
\rightskip 2pc
\noindent{\ninepoint\sl \baselineskip=8pt {\bf Fig.3}: {\rm Momentum
modes, with energy $n/L$ get exchanged with winding modes with
energy $mL$ under mirror symmetry $L\rightarrow 1/L$.}}

There is one context in which a similar duality is already well known
mathematically:  Consider a $U(1)$ bundle on a circle.  Then the choice
of the bundle (i.e. the choice of the holonomy of $U(1)$ around the circle) is
equivalent to the choice
of a point on the dual circle.  This also turns out to have a very important
physical analog \ref\polc{J. Polchinski,
Phys. Rev. Lett. {\bf 75} (1995) 4724.}. If we consider open strings, in
addition to closed strings,
we would be considering Riemann surfaces with boundaries.  In such a case
in addition to specifying the target geometry $M$ where the
closed strings are mapped to, we have to specify where
the boundaries are mapped to.  In general they could map to some
subspaces of $M$ of various dimensions $p$.  Such a p-dimensional
subspace of $M$ is called a $p-brane$ or $Dp-brane$ ($D$ signifying
the fact that the maps from the Riemann surface have {\it Dirichlet} conditions
in codimension $p$, and ``brane'' generalizing the terminology of membranes
which are 2-branes, to the higher dimensional objects).  Moreover it turns
out that a $Dp$-brane will carry a $U(1)$ gauge field
and so can be viewed as a sheaf in $M$.  Physically a $Dp$-brane corresponds
to some charged matter localized in a $p$-dimensional subspace of $M$.
{}From the string viewpoint
D-branes are regions where an open string can end on (Fig. 4).

\bigskip
%\midinsert
\epsfxsize 2.truein
\epsfysize 2.truein
\centerline{\epsfxsize 2.truein \epsfysize 2.truein\epsfbox{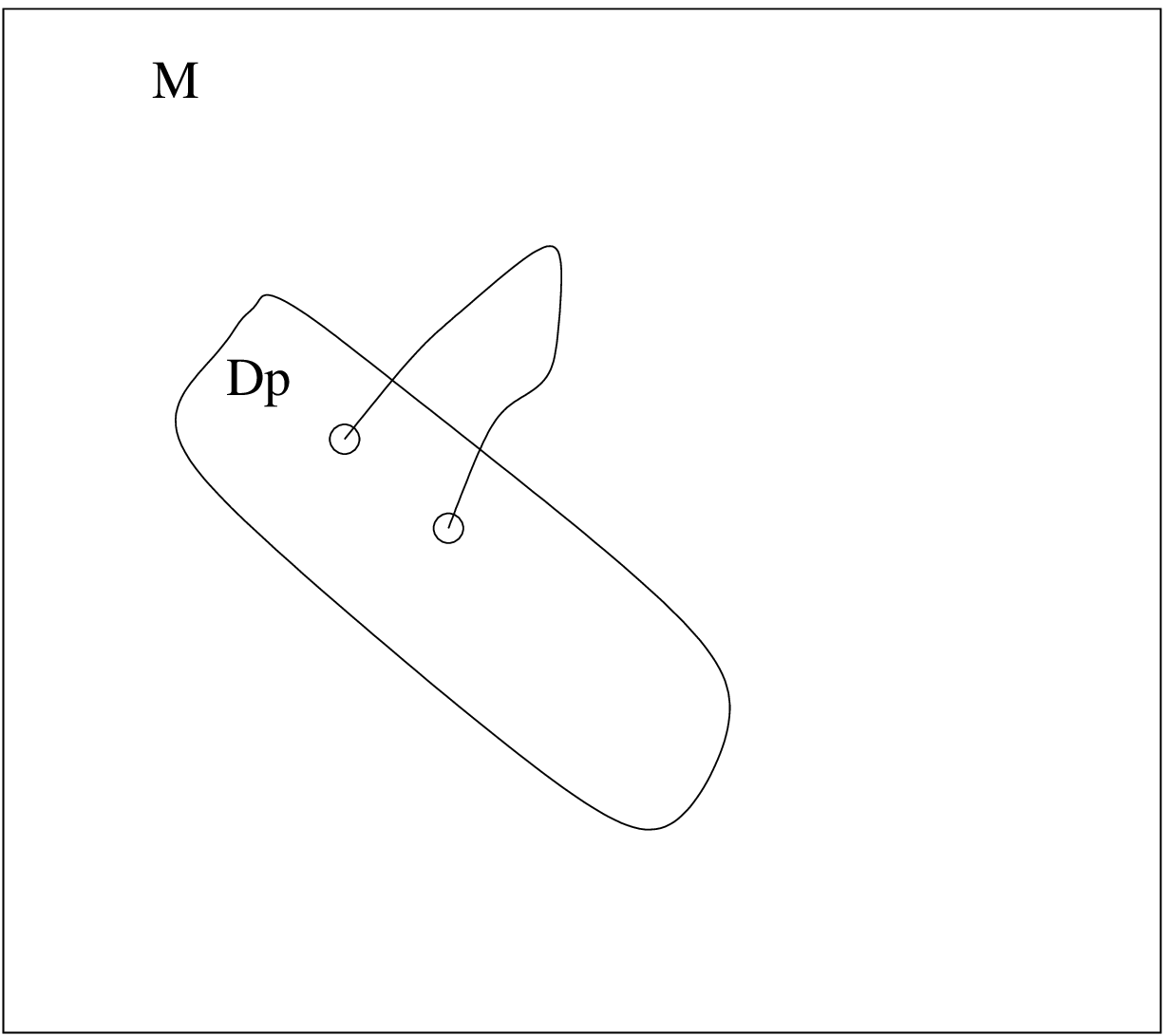}}
\leftskip 2pc
\rightskip 2pc
\noindent{\ninepoint\sl \baselineskip=8pt {\bf Fig.4}: {\rm A $Dp$ brane
is a subspace of the target manifold $M$ where a string can end on.}}

Returning to the case of a circle, if we consider a $D1$ brane which
includes the entire circle of circumference $L$, we can ask what
happens under mirror symmetry to the D-brane.  The answer is that
it gets transformed to a $D0$ brane on the mirror.  This is in accord
with the mathematical fact mentioned before (where the holonomy of a $U(1)$
bundle gets transformed to the choice of a point on the dual circle).  This
has also a natural generalization to the case where we consider $N$ D1 branes
wrapping the $S^1$ which in physics leads to a $U(N)$ bundle
on $S^1$ and choosing a flat $U(N)$ connection on $S^1$ amounts to choosing
$N$ points on the dual circle, i.e. it is transformed to $N$ $D0$ branes on the
mirror.

It is natural to ask how mirror symmetry extends in cases
where the target manifold is more complicated than $S^1$.  One simple example
consists of taking a d-dimensional torus $T^d=(S^1)^d$ and doing inversion on
each of
the $S^1$'s.  The action of this on the $Dp$ branes, viewed as
subspaces $T^p\subset T^d$
is also clear where they get transformed to a dual $T^{*d-p}\subset T^{*d}$.
 However for more interesting examples
we need the following idea \foot{The presentation here of the mirror
symmetry for more complicated target spaces does not follow the historical
order
of its discovery.
Mirror symmetry was first conjectured to exist for Calabi-Yau
manifolds in \ref\lvw{W. Lerche, C. Vafa and N. P.
Warner, Nucl.Phys. {\bf B324} (1989) 427.}\ref\dix{ L. Dixon in Proc. of the
1987 {ICTP Summer Workshop in
High Energy Physics and Cosmology}, Trieste.}, with the concrete examples
being found in \ref\gp{B. R. Greene and M. R.
Plesser, Nucl. Phys. {\bf B338} (1990) 15.}\ followed by a concrete
application to counting holomorphic curves in \ref\can{P. Candelas, Xenia C. de
la Ossa,
Paul S. Green and L. Parkes, Nucl. Phys. {\bf B359} (1991) 21}.
The construction of mirror pairs was systematized by \ref\bat{V. Batyrev,
J. Alg. Geom. {\bf 3} (1994) 493.}.
The presentation here follows the approach in \ref\syz{A. Strominger,
S.-T. Yau and E. Zaslow, Nucl. Phys. {\bf B479} (1996) 243.}\
developed further in \ref\lev{N.C. Leung and C. Vafa,
Adv. Theor. Math. Phys. {\bf 2} (1998) 91.}\ which explains
the construction of \bat\ from this viewpoint. }.

\subsec{The Adiabatic Principle}
Consider a family of flat d-dimensional tori $T^d$ {\it varying slowly}, i.e.
{\it adiabatically} over
some base space $B$.  Consider the total space $M_1$ over
$B$ with $T^d$ as the fiber.  Consider another space consisting of the same
base space $B$, where over each point we replace the fiber $T^d$ with the
mirror torus where all lengths are inverted. Call the total space $M_2$.
Then it is natural to believe that the spaces $M_1$ and $M_2$ are mirror
to one another.  However the interesting examples arise when the assumption
of adiabaticity is violated over some subspaces of $B$.  For example
the $T^d$ may degenerate over some loci. If the category
of objects we are dealing with is sufficiently nice one may hope that
the mirror property will continue to hold.  One nice category\foot{
There may well be other categories, such as the category
of manifolds of $Sp(n)$, $Spin(7)$ or $G_2$ holonomy.}
seems to be when the base $B$ is also $d$-dimensional
and the total space is a Calabi-Yau d-fold (a K\"ahler manifold of
complex dimension $d$ whose bundle of holomorphic $d$-forms is
trivial) where the fibers
$T^d$ are viewed as Lagrangian submanifolds relative to the
K\"ahler form.  In fact the non-trivial data specifying
the geometry of the Calabi-Yau is precisely how the degeneration
of $T^d$ over $B$ takes place.  This construction
corresponds to describing a hypersurface in a toric variety, in a degenerate
limit.
In a singular limit the Calabi-Yau
may be viewed as a $T^d$ fiber space over the base being a boundary
of some simplex (in the sense of toric geometry), where $T^d$ degenerates
to $T^k$ over $d-k$ dimensional subspaces of $B$.   The data defining
the mirror, after suitably rescaling the metric on $B$ looks like
the dual geometry where the regions where the $T^d$ shrinks to $T^k$
is replaced by the dual $k$-dimensional subspaces where the $T^{d-k}\subset
T^d$
shrinks and the dual survives, this being consistent with the
small/large radius exchange (Fig. 5).  This gives what is known
as Batyrev's construction of mirror pairs using the toric description.

\bigskip
%\midinsert
\epsfxsize 2.truein
\epsfysize 2.truein
\centerline{\epsfxsize 2.truein \epsfysize 2.truein\epsfbox{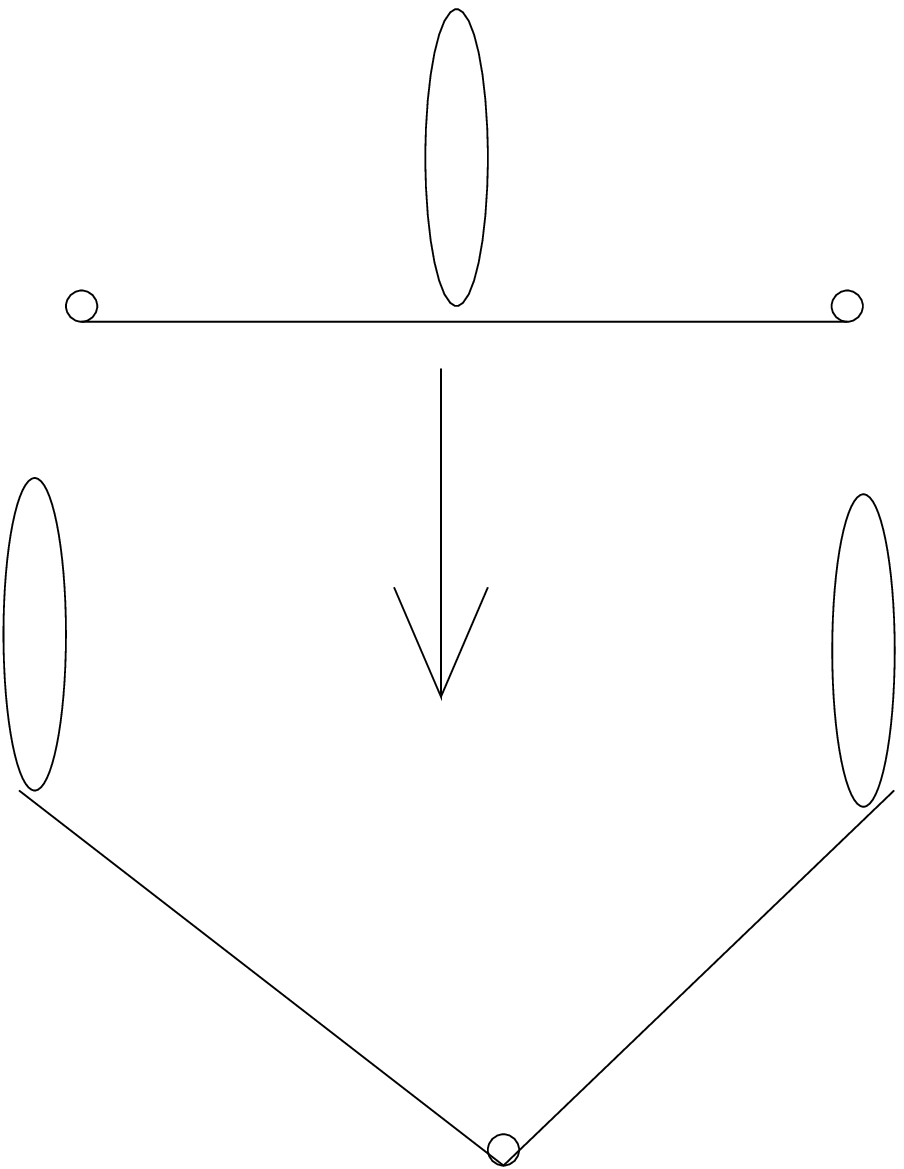}}
\leftskip 2pc
\rightskip 2pc
\noindent{\ninepoint\sl \baselineskip=8pt {\bf Fig.5}: {\rm An application
of inversion duality of tori when tori are varying leads to an explanation
of mirror symmetry in more complicated examples.}}

\subsec{K\"ahler-Complex Deformation Exchange}
It would be nice to examine some of the consequences of the existence of
mirror geometries. To get a feeling for this it is useful
to start at the level of $S^1$ fibered trivially over $B=S^1$.  This is
a simple case, as a constant fibration admits the flat metric.
Let $R_f,R_b$ denote the radii of the fiber and a section
respectively.  Note that the complex structure (shape) of the
torus is determined by
$$C=R_b/R_f$$
and its K\"ahler class (size) is determined by
$$K=R_b R_f$$
Now if we do mirror transform on the fiber $S^1$ it again leads to a torus.
However since $R_f\rightarrow 1/R_f$ but $R_b\rightarrow R_b$
the parameters controlling the complex and K\"ahler deformations get exchanged:
$$C\leftrightarrow K \qquad under \quad mirror \quad transform$$
This turns out to be the general feature of mirror symmetry for Calabi-Yau
manifolds, and the K\"ahler and complex structures always get exchanged.
In the case of Calabi-Yau manifold of complex dimension
$d$ the number of complex moduli
is determined by $h^{1,d-1}$
(where $h^{p,q}$ denotes the dimension of the
cohomology of $p$-holomorphic and $q$ anti-holomorphic forms). Thus if $M$ and
$W$
are mirror Calabi-Yau manifolds we learn in particular that
$$h^{1,1}(M)=h^{1,d-1}(W)\qquad h^{1,d-1}(M)=h^{1,1}(W).$$
This in particular implies that the topology of the manifold and the mirror
will in general be very different.  In fact it turns out that $h^{p,q}(M)=
h^{p,d-q}(W)$ for all $p,q$.
  Moreover, as mentioned before,
the parameter controlling quantum corrections is the K\"ahler class of the
Calabi-Yau, which gets transformed under mirror transform to complex
deformation
parameter of the mirror.  Thus the question of quantum corrections for one
manifold get transformed to the question involving the variation of complex
structure
on the other, which is classical.  This leads to some very non-trivial
implications of mirror symmetry.

The most concrete prediction this leads to is
to the question of counting the ``number'' of holomorphic curves
mapped from a Riemann surface of genus $g$ to the threefold.  For example
the intersection numbers of cycles in the Calabi-Yau receives a quantum
correction
coming from holomorphic curves (recall this is natural from the
string theory viewpoint, where the worldsheet is a Riemann surface)
(Fig. 6).  This ``quantum intersection theory'' for triple intersections
allows, in addition to the classical intersection, the possibility
that the three cycles meet a holomorphic curve weighted by
the quantum deformation
parameter $q=e^{-A}$ where
$A$ is the
area of the holomorphic curve\foot{The fact that classical cohomology
ring is deformed by instantons and gives rise to a quantum cohomology ring
was pointed out in \lvw. The precise definition of this deformation
was given in \ref\witqc{E. Witten, Nucl. Phys. {\bf B340} (1990) 281.}.}.

\bigskip
%\midinsert
\epsfxsize 2.truein
\epsfysize 2.truein
\centerline{\epsfxsize 4.truein \epsfysize 2.truein\epsfbox{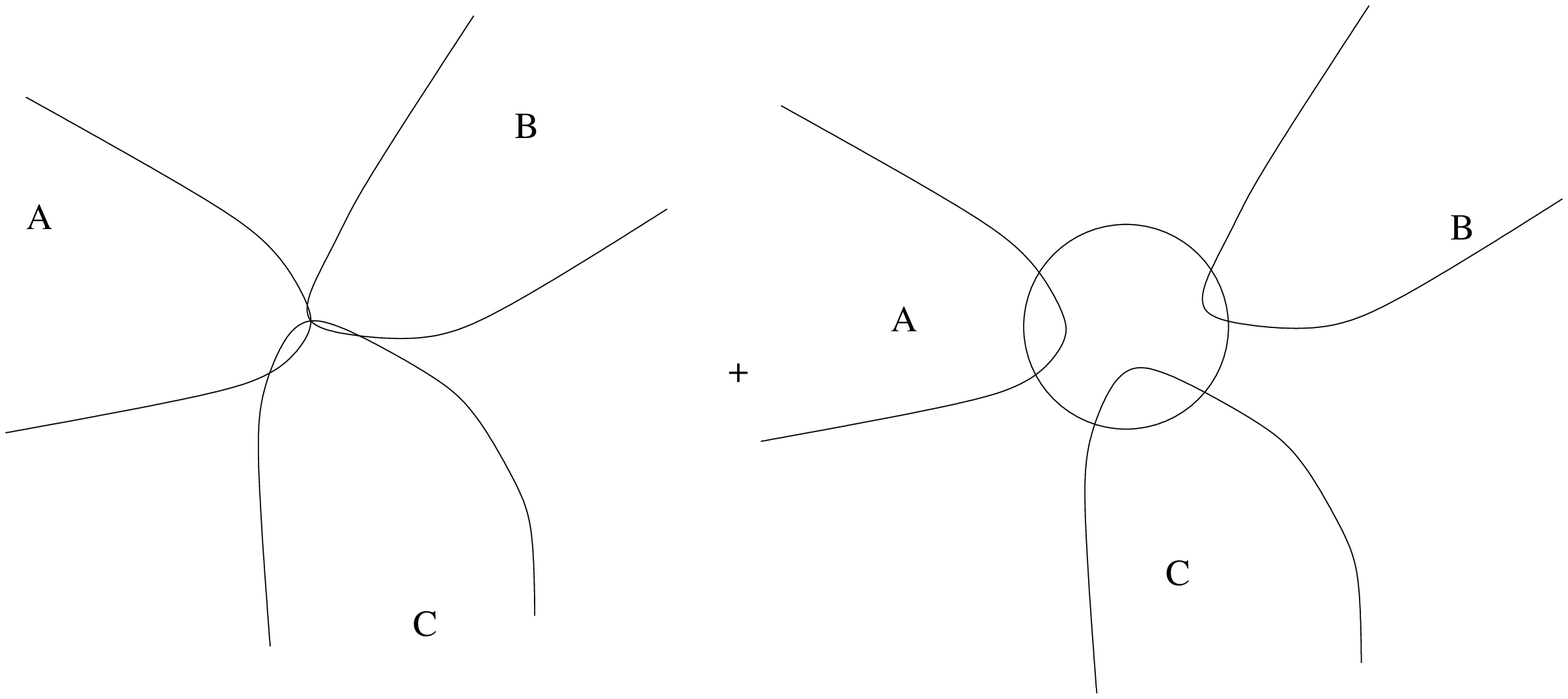}}
\leftskip 2pc
\rightskip 2pc
\noindent{\ninepoint\sl \baselineskip=8pt {\bf Fig.6}: {\rm Quantum
intersection of three cycles $A,B,C$ in addition to the classical
piece has corrections where $A,B,C$ meet on a holomorphic rational
curve.}}

This very
difficult mathematical problem, i.e. counting holmorphic curves in Calabi-Yau
manifolds, gets transformed on the mirror to a question involving the variation
of
Hodge structures (in this case
it is the study
of how the middle dimensional $H^{p,d-p}$ cohomology elements vary as we vary
the
complex structure on the mirror).  This is a well studied mathematical subject
\foot{To be precise, the counting of genus 0
curves gets transformed to this question.  The higher genus version
gets transformed to a quantum version of variation of Hodge structure known
as {\it Kodaira-Spencer theory of gravity} which
is only slightly more complicated.}.  The genus 0 version of the prediction
    has been made rigorous recently
\ref\giv{A. Givental,
  Internat. Math. Res. Notices (1996),
613 .}\ref\yet{B.H. Lian,
K. Liu and S.-T. Yau, Asian J. Math. {\bf 1} (1997), 729.}.  The higher
genus version \ref\bcov{M. Bershadsky, S. Cecotti, H. Ooguri and C. Vafa,
    Comm. Math. Phys. {\bf 165} (1994) 311.}\
has
not been proven yet (except in some special cases), but there is little doubt
that it is generally valid.

\subsec{Extension to Bundles}
It is clear from the discussion of D-branes in the context of circles
that we can extend mirror symmetry to Calabi-Yau manifolds with bundles.  In
particular
let $c\in \oplus_p H^{p,p}(M)$ denote the chern class of
a holomorphic vector bundle on Calabi-Yau manifold $M$.  Represent
this by a collection of Poincar\'e dual holomorphic cycles.  Consider
D-branes wrapped over them.  This is a D-brane made up of various
even dimensional branes.  Each $(p,p)$ cycle projects to
 a $p$ real dimensional
subspace of $B$ with typical fiber a $p$ dimensional subtorus.  On the mirror,
the $p$
dimensional subspace of $T^d$ gets
transformed to the dual torus $T^{d-p}$.  Thus on the mirror Calabi-Yau,
the whole bundle representated by the collection of D-branes
is mirror to a submanifold $C$ of real dimension $d$.\foot{
This leads to a new application of mirror symmetry:  For example consider a
rational elliptic surface inside a 3-fold.  Then the study of rank $N$ stable
bundles
on it gets transformed to the study of spectral curves on the dual
rational elliptic surface (by viewing the bundle as D4 brane
wrapped the rational elliptic surface and doing mirror symmetry
along $T^2$ fiber)\ref\fmw{R. Friedman, J.W. Morgan and E. Witten,
{\it Vector Bundles over Elliptic Fibrations} alg-geom/9709029.}\ref\berpa{M.
Bershadsky, A. Johansen, T. Pantev and V. Sadov, Nucl. Phys. {\bf B505} (1997)
165.}.
 The Euler class of the moduli space can be computed using mirror
symmetry techniques \ref\mnvw{J. Minahan, D. Nemeschansky, C. Vafa and
N.P. Warner, {\it E-strings and N=4 Topological Yang-Mills Theories},
hep-th/9802168.}\ (this prediction has been recently
confirmed for the rank 2 case \ref\yosh{K. Yoshioka, {\it
Euler characteristics of SU(2) instanton moduli spaces on rational
elliptic surfaces}, math.AG/9805003.}).}
The condition that the original bundle be holomorphic translates
to the condition that $C$ is Lagrangian relative to the K\"ahler form on the
mirror.  If we further impose that the original bundle be stable, this
translates to the cycle $C$ being of minimal area.  This extension of mirror
symmetry to include bundles conjectured in  \ref\cva{C. Vafa, {\it
Extending Mirror Conjecture to Calabi-Yau with Bundles}, hep-th/9804131.}
(see also related works \ref\kon{M. Kontsevich, Proceedings of the 1994 {\bf
ICM I}, Birk\"auser, Z\"urich, 1995, p.120; alg-geom/9411018.}\ref\fuk{K.
Fukaya, The Proceedings of the 1993 GARC Workshop on Geometry and
Topology, H.J. Kim, ed., Seoul National University.}\ref\zas{A. Polishchuk and
E. Zaslow, {\it Categorical Mirror Symmetry: The Elliptic Case},
math.AG/9801119.}\ref\tyu{A.N. Tyurin, {\it Non-Abelian
analogues of Abel's Theorem}, I.C.T.P. preprint (1997).}) has only recently
been made and checks on its prediction are underway.  It makes certain
predictions for the enumerative geometry of holomorphic maps from Riemann
surfaces with boundaries being mapped holomorphically
to Calabi-Yau, with boundaries being mapped to Lagrangian
cycles on it.\foot{For this to make sense beyond Disc
one should restrict to the category of stable
bundles on one side and minimal Lagrangian
submanifolds on the mirror.}.  For example the Ray-Singer
Torsion associated to the bundle $V$ is transformed to counting
holomorphic maps from the
annulus to the Calabi-Yau whose boundary is on the mirror minimal cycle.

\newsec{Physical Interpretation of Geometric Singularities}
One of the remarkable aspects of string theory is the existence
of a few different types of consistent theories (5 in 10 dimensions
and one in 11 dimensions) which are dual to one another.  This is known
as S-duality.
For example, Type IIA strings in
a 10 dimensional space having a $K3$ fibration
($K3$ being a Calabi-Yau manifold of complex dimension 2)
is dual to heterotic strings in a space admitting a $T^4$ fibration.
This is
very surprising because in particular the two string theories and the two
target spaces
look very different. Moreover on the heterotic side one has to choose
flat bundles of rank 16.  Moreover as we change the
size of the $T^4$ and the choice of the flat bundle (and some choice
of a constant field belonging to $H^2(T^4))$ one can get various
different gauge groups.  For example one can obtain $SU(N),SO(2N)$
(for small enough $N$) and $E_{6,7,8}$.  The question is how all
this is reflected on the $K3$ geometry?   It is well known that
$K3$ can have singularities corresponding to contracting 2 spheres.
Moreover the intersection matrix of the contracting 2 spheres is
given by the Cartan matrix of the A-D-E groups.  The appearance
of the Dynkin structure for the $K3$ singularities appears mathematically
as purely ``accidental''.  However this accident gets explained in this
duality context:
One identifies the singular $K3$ geometries with A-D-E
singularities with the points on the heterotic side with enhanced
A-D-E gauge symmetry.
The physical explanation of enhanced symmetries on the $K3$ side has
to do with the existence of D2 branes, which can wrap around the
contracting 2-cycles, and give rise to massless particles.  The wrapped D2
branes encode in a beautiful way
the connection of the bundle anticipated from the
heterotic dual (Fig. 7).  Thus the non-abelian enhancement of gauge symmetry on
heterotic side
is transformed to appearance
of geometric singularities on the type IIA side.

\bigskip
%\midinsert
\epsfxsize 2.truein
\epsfysize 2.truein
\centerline{\epsfxsize 4.truein \epsfysize 2.truein\epsfbox{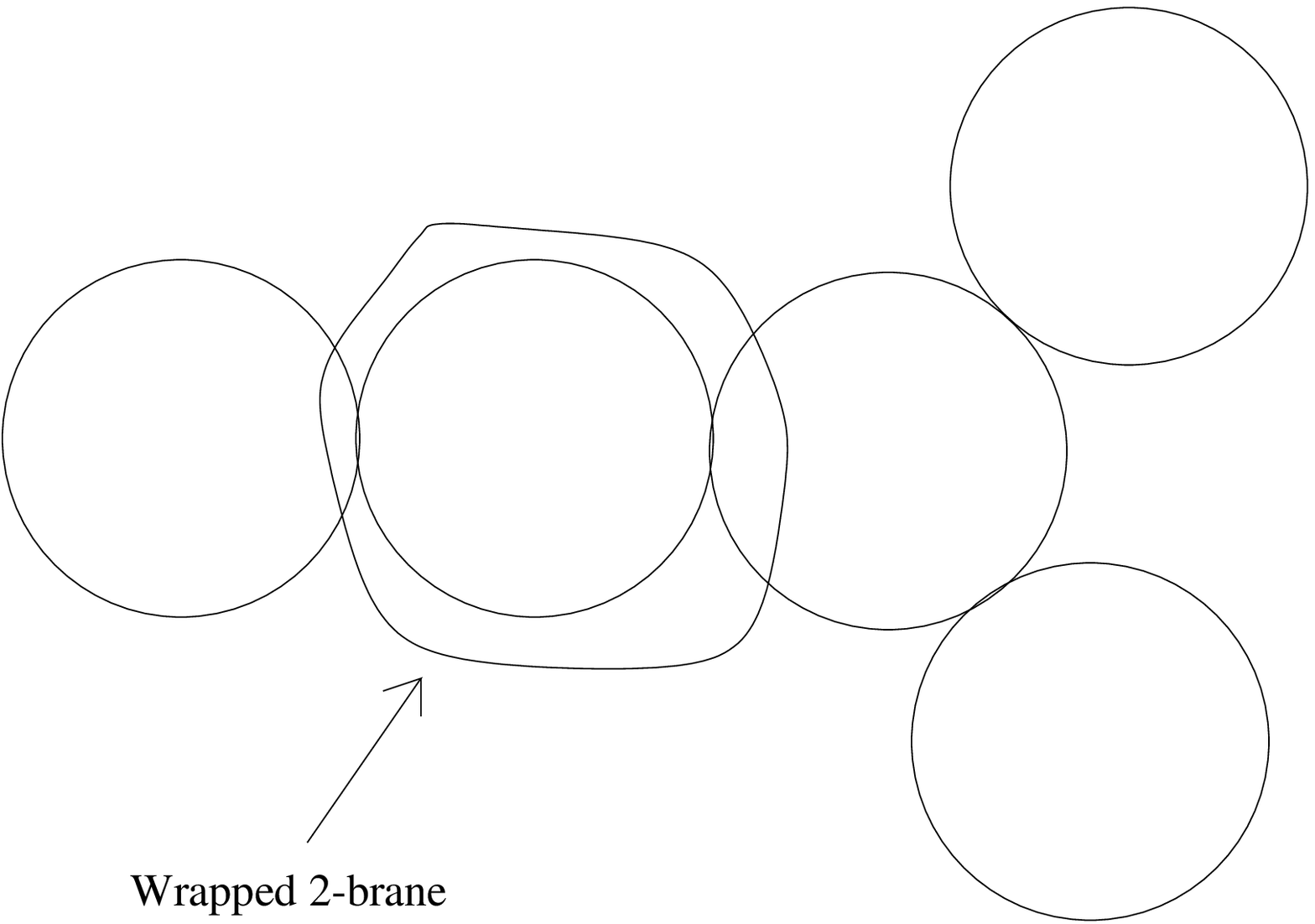}}
\leftskip 2pc
\rightskip 2pc
\noindent{\ninepoint\sl \baselineskip=8pt {\bf Fig.7}: {\rm A wrapped
D2-brane over a sphere of blown up A-D-E- singularity is
the origin of gauge symmetry enhancement when the spheres shrink.}}

Similar considerations suggest interesting
physical interpretations whenever one has geometric singularities.  For example
if one considers a Calabi-Yau 3-fold, one has sometimes contracting $S^3$'s.
In  this context there are two ways to get rid of the singularity.  One either
deforms the polynomial equations defining the manifold
(which effectively
gives a finite size to the contracting $S^3$'s) or replaces
the singular point by a higher dimensional geometry (in this case $S^2$'s)
which is known as
 blowing up the singularities,  changing the geometry
of the 3-fold in the process.  The singular manifold can thus be viewed as
belonging to two distinct families of Calabi-Yau manifolds.   The physical
interpretation of this is
that there are two ways to get rid of the extra massless fields, one is by
preserving a
$U(1)^k$ gauge symmetry which is called the ``Coulomb branch'' (corresponding
in type IIA string
to blowing up $S^2$'s) the other is
going to the ``Higgs branch''
(which corresponds to making $S^3$'s have finite volume)\ref\strom{A.
Strominger,
Nucl. Phys. {\bf B451} (1995) 96.}\ref\gms{B. Greene, D.R. Morrison
and A. Strominger, Nucl. Phys. {\bf B451} (1995) 109.}.

One can use these ideas to construct the geometric versions
of quantum field theories with desired properties.  This is called
{\it geometric engineering of quantum field theories}.  For example,
if we have a shrinking ${\bf CP^1}$ in $K3$ we already mentioned
that this gives rise to $SU(2)$ gauge symmetry.
If we fiber this over a complex curve, depending on what curve
we choose we get different theories in the 4 left-over dimensions.  For
example if we consider the simple product with $T^2$, then we obtain a
theory in four dimensions with $N=4$ supersymmetric $SU(2)$
Yang-Mills theory.  Moreover the coupling constant of the gauge theory
$1/g^2$ (which appears in the action in 4 dimensions in the form
${1\over g^2}Tr F\wedge *F$)
gets identified with the volume of $T^2$.  As discussed before
string theory has volume inversion symmetry for $T^2$.  This implies,
therefore, that $N=4$ Yang-Mills should have $g\rightarrow 1/g$
inversion symmetry as well.  This in fact was anticipated
    long ago \ref\mo{C. Montonen and D. Olive, Phys. Lett. {\bf B72} (1977)
117.}.  This duality has interesting
consequences for four-manifolds: Consider taking as the four left-over
dimensions a smooth four manifold $K$.  Then the (topological) partition
function
of $N=4$
Yang-Mills is given by
$$F_{G,K}(q)=\sum_{instantons}q^n \chi({\cal M}_n)$$
where $q={\rm exp}(-1/g^2)$ and $\chi({\cal M}_n)$ denotes the
Euler characteristic of the moduli space of instantons of gauge
group $G$ (in the case at hand $G=SU(2)$) with instanton number $n$ on $K$.
The duality just discussed implies that this is a modular form (after
shifting by an overall coefficient $q^a$ for some constant $a$).
This has been tested in some cases (see \ref\vw{C. Vafa and E. Witten,
Nucl. Phys. {\bf B431} (1994) 3.}\ and references therein).
This modular form is a smooth invariant of $K$, for each group $G$.\foot{
The subgroup of $SL(2,{\bf Z})$ for which this is a modular form depends on
$G$.}

If we fiber the $A_1$ singularity instead of $T^2$ over a ${\bf CP^1}$ we
obtain
an $N=2$ supersymmetric
gauge theory in 4 dimensions with $SU(2)$ gauge symmetry.   If different
singularities exist over different curves which intersect
(what is sometimes called {\it colliding singularities}) we typically
get ``matter'' in the physical language transforming according to
a representation of the product of the two groups (Fig. 8) \ref\kav{S. Katz
and C. Vafa, Nucl. Phys. {\bf B497} (1997) 146 \semi S. Katz,
A. Klemm and C. Vafa, Nucl. Phys. {\bf B497} (1997) 173\semi
S. Katz, P. Mayr and C. Vafa, Adv. Theor. Math. Phys. {\bf 1} (1998) 53.}.

\bigskip
%\midinsert
\epsfxsize 2.truein
\epsfysize 2.truein
\centerline{\epsfxsize 4.truein \epsfysize 2.truein\epsfbox{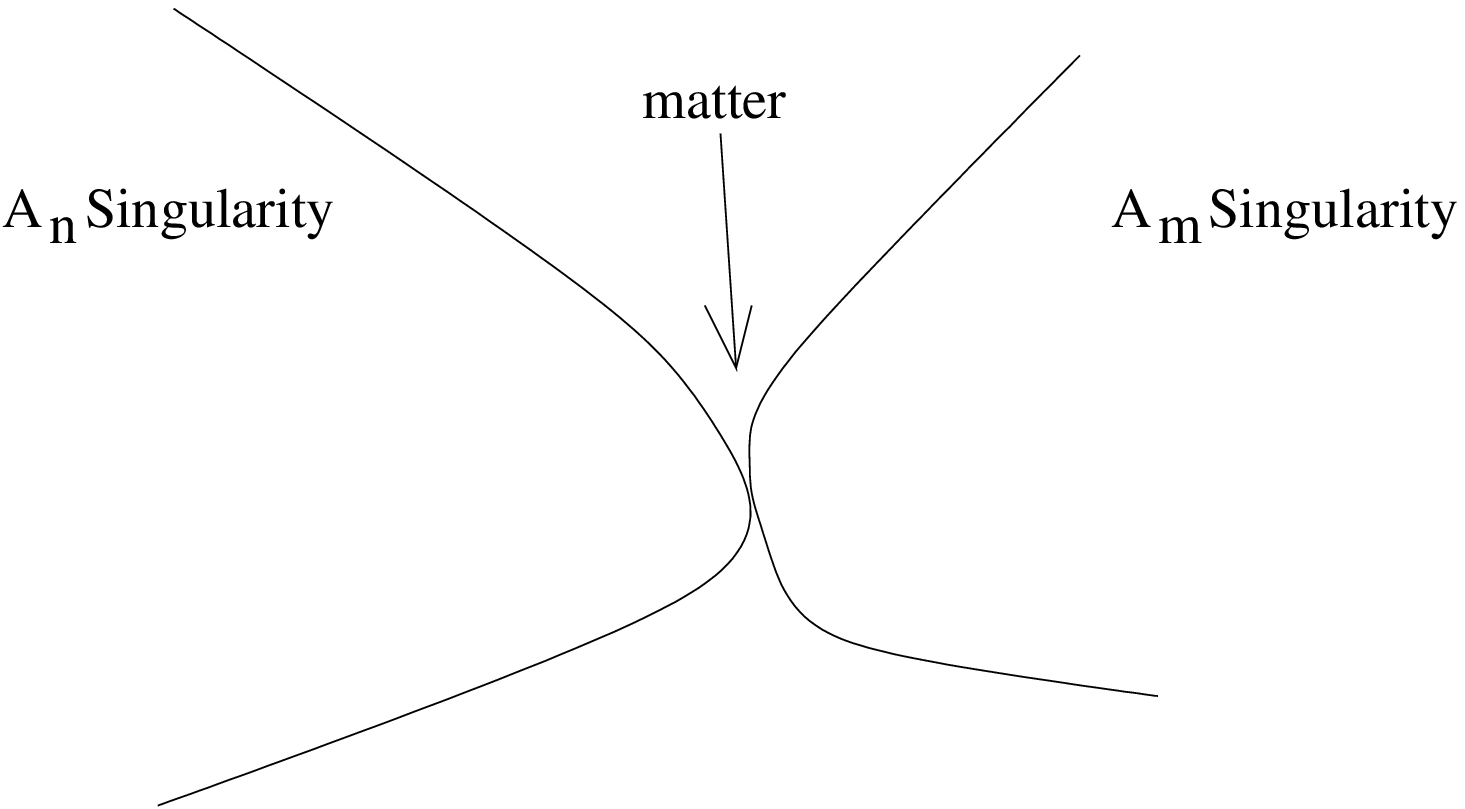}}
\leftskip 2pc
\rightskip 2pc
\noindent{\ninepoint\sl \baselineskip=8pt {\bf Fig.8}: {\rm Matter
arises where two loci of singularities intersect.  The matter
is localized at the intersection.}}

This geometric
construction of quantum field theories allows us to have
a new viewpoint in solving aspects of them. For example
consider the $N=2$ supersymmetric $SU(2)$ gauge theory in 4 dimensions.
As just mentioned this can be viewed as fibering a contracting ${\bf CP}^1$
over a base ${\bf CP}^1$.  The instantons of this theory
in four dimensions, which are relevant to questions
involving Donaldson invariants of four manifolds,
correspond to holomorphic curves mapped to a Calabi-Yau 3-fold
whose local geometry is a line bundle over a ${\bf CP}^1$ fibered
over ${\bf CP^1}$.   In particular the instanton class in four dimension
gets identified with the number of times the curve gets wrapped around
the base ${\bf CP}^1$.
 These can be counted thanks to mirror symmetry
discussed before.  Thus Donaldson invariants \ref\don{S.K. Donaldson,
Proc. London. Math. Soc. {\bf 50} (1985) 1.}\
through this geometric
construction and by an applications of mirror symmetry can be reduced
to Seiberg-Witten invariants \ref\seibw{N. Seiberg and
E. Witten, Nucl. Phys. {\bf B431} (1994) 484.}\ref\witmoor{G. Moore
and E. Witten, {\it Integration over the u-plane in Donaldson
theory}, hep-th/9709193.}.

Sometimes the physics of the singularities are unconventional.
For example when a 4-cycle (say a ${\bf CP}^2$) shrinks in a Calabi-Yau
threefold, it gives rise to very interesting unconventional
new physical theories
which were not anticipated! This is thus a great source of insight into
new physics.  In particular what types of singularities occur as well
as what are the ways to resolve them will be of extreme importance for
unravelling aspects of this new physics.
It is tempting to speculate that these singularities may also lead
to new invariants for four manifolds.

\newsec{Black Holes and Minimal Cycles}
Black holes are solutions to the Einstein equations which represent
matter with sufficient concentration in some region.\foot{
The following discussion is somewhat oversimplified to make the essential point
more clear.}
Consider a $d$ dimensional spacetime.
The idealized version of a black hole would correspond to a spherically
symmetric
distribution of possibly charged matter.  This would correspond
to solving Euler-Lagrange equations for the action of the form (suppressing all
constants)
$$S=\int (R+\sum_i F_i\wedge *F_i)$$
where R denotes the scalar curvature of the metric and $F_i$ denote the
curvature of
some $U(1)^k$ gauge fields.  One solves these equation with the assumption
of spherical symmetry with some asymptotic condition imposed on the
metric which corresponds to a total mass $M$ black hole and on the gauge fields
with charge $Q_i=\int_{S^{d-2}}*F_i$.\foot{If d=4
we can also consider having magnetic charges $M_i=\int_{S^{2}}F_i$.}

Black holes have a causal structure which separates it into two parts
by a ``horizon'' $H=S^{d-2}$, for which the future light cone of points
inside the sphere does not include exterior points (Fig. 9).

\bigskip
%\midinsert
\epsfxsize 2.truein
\epsfysize 2.truein
\centerline{\epsfxsize 2.truein \epsfysize 2.truein\epsfbox{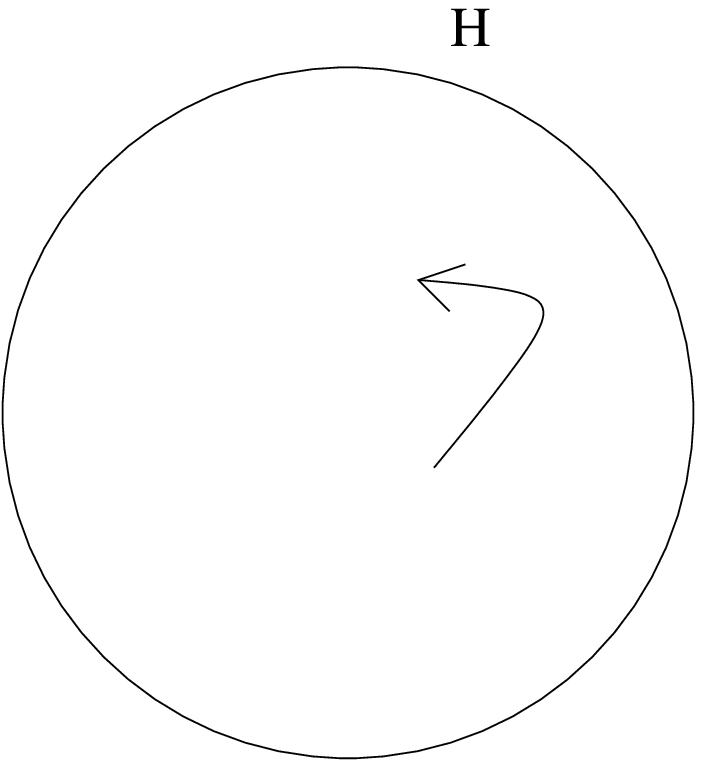}}
\leftskip 2pc
\rightskip 2pc
\noindent{\ninepoint\sl \baselineskip=8pt {\bf Fig.9}: {\rm
{}From the regions
interior to the horizon no light can come out.}}

 By some semiclassical arguments
one expects that black hole carry entropy $S$, which is the logarithm
of the number of its states, is given by
$$S={A(H)\over 4}$$
where $A(H)$ denotes the $d-2$ dimensional ``area'' of the
horizon $H$.
For the black hole solution to make physical sense one finds a lower
bound on mass for a fixed set of charges $Q_i$, namely $M^2\geq \sum_i Q_i^2$.
Physically what will happen is that if the mass is above this bound
the black hole radiates and loses mass until it reaches this bound, at which
point it becomes a stable stationary state.  These are known as {\it
extremal black holes}.
The entropy, which is
defined as a quarter of the horizon area now becomes
$$S=c_d M^{{d-2}\over d-3}$$
where $c_d$ is some universal constant, depending on $d$.
It has been a challenge of quantum gravity to explain the microscopic
origin
of this entropy, i.e. what counting do we do to get this entropy.

In string theory, for large enough charges $Q_i$,
the charged black holes are realized as
branes wrapped around cycles of the Calabi-Yau, and the condition for
extremality of the black hole is that the corresponding cycle be
minimal in the given class.  Thus the charge lattice corresponds
to $H_*(M)$ where the target space is $R^d\times M$.\foot{
The homology dimensions which are allowed charges
correspond to the allowed dimensions of the branes in
the corresponding theory.}.   Thus
the question of black hole entropy gets transformed to counting of the
``number'' of minimal submanifolds for a fixed class $Q\in H_*(M)$.
In case there are moduli for such cycles, what is meant by
the ``number'' is the number of cohomology elements of the moduli space.
The non-minimal surfaces correspond to non-extremal black holes which
``decay'' to the extremal ones.

I will now discuss one concrete example to illustrate how the counting
works.  Consider the 11 dimensional
supergravity theory (``M-theory'') on target space $R^5\times T^6$ (which
is closely related to type IIA on $R^4\times T^6$), which I will use
to count the number of black holes in 5 dimensions, with charges
given by an element in $H_2(T^6,{\bf Z})$ (this is related
to black hole count in \ref\stomv{A. Strominger
and C. Vafa, Phys. Lett. {\bf B379} (1996) 99.}).  Let us think
of $T^6=(T^2)^{3}$ and consider the 2-class
of each $T^2$ being represented by $e_i$ where $i=1,2,3$. Let us consider
an extremal black hole made of 2-branes whose class is $Ne_1+M e_2+P e_3$.
We will consider the regime of parameters where $N>>M,P>>1$.  Let $\Sigma$
denote a holomorphic curve in the class $[\Sigma ]=Me_2+Pe_3$ (being
holomorphic
guarantees being minimal in that class).  To construct a 2-surface
in the class $Ne_1+Me_2+P e_3$ we choose $N$ points on $\Sigma$ and attach
a copy of the first $T^2$ on each of those points (Fig. 10).

\bigskip
%\midinsert
\epsfxsize 2.truein
\epsfysize 2.truein
\centerline{\epsfxsize 4.truein \epsfysize 2.truein\epsfbox{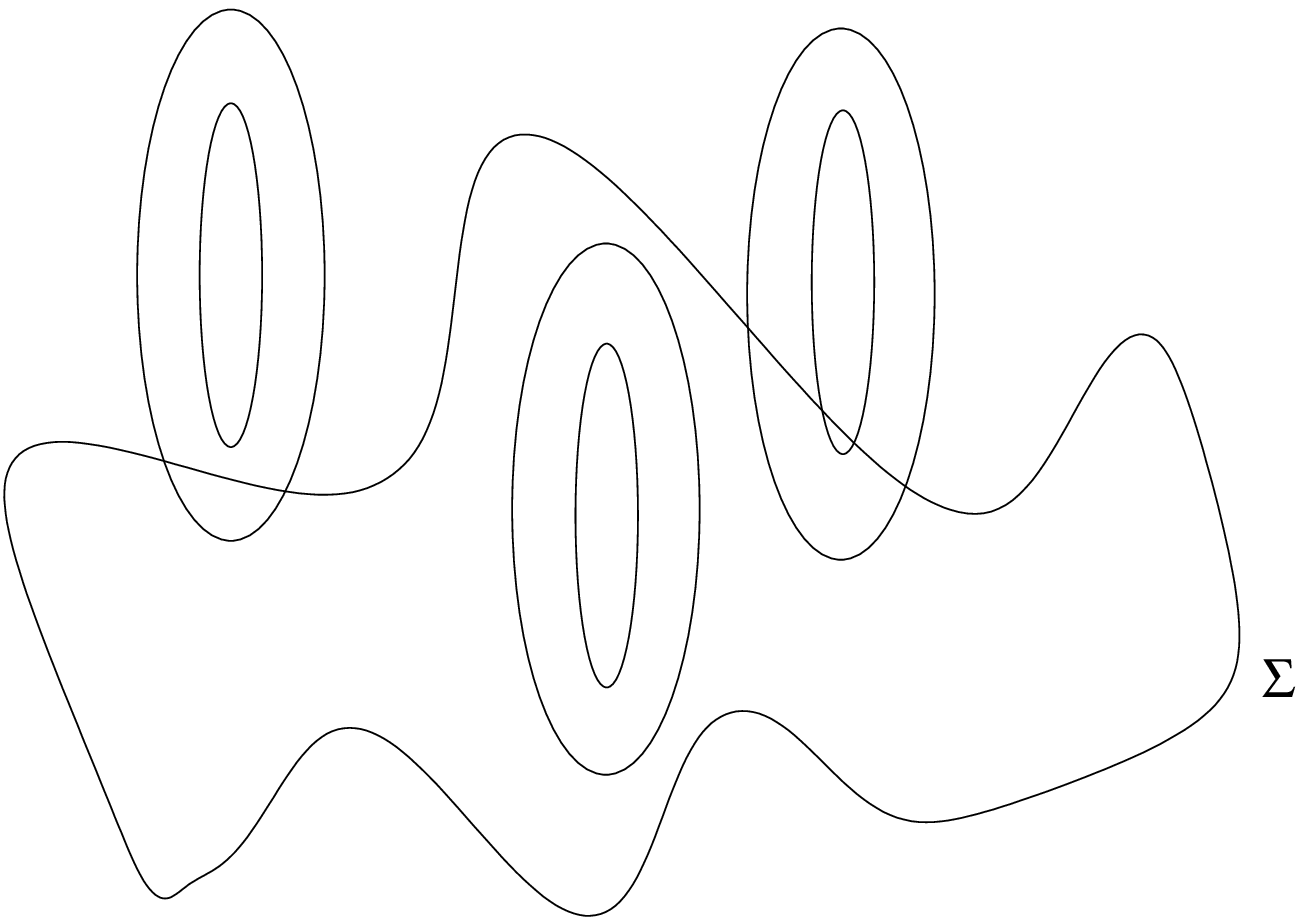}}
\leftskip 2pc
\rightskip 2pc
\noindent{\ninepoint\sl \baselineskip=8pt {\bf Fig.10}: {\rm A 2-brane
constructed out of $\Sigma$ and the attachment of $N$ copies of $T^2$
at $N$ points.}}

  This gives rise
to a degenerate minimal 2-cycle.  The moduli of this $D2$ brane will in
addition
correspond to choosing a flat connection on it, which for each $T^2$
corresponds to choosing a point on the dual $T^2$.  Thus this surface
together with the choice of a flat connection is specified by $N$ points
in ${\hat T}^2\times \Sigma$ where ${\hat T}^2$ denotes the dual torus.
Of course the choice of $N$ points has no ordering so that the moduli
space of this minimal cycle, for a {\it fixed } $\Sigma$
is given by
$${\cal M}_N={\rm Sym}^N (T^2\times \Sigma)$$
Since we are interested in the regime where $N$ is much
larger than the other two parameters, we can treat
$\Sigma$ as fixed (i.e. the moduli degrees of freedom coming
from it are negligible in comparison).  We are thus interested
in the growth for the cohomology of ${\cal M}_N$ for large $N$.
This space is singular and this cohomology should be understood
in the sense of the Hilbert Scheme.  The answer is well known \ref\Gott{L.
G\"ottsche, Math. Ann. {\bf 286} (1990) 193.}\ref\hirz{F.
Hirzebruch and T. H\"ofer, Math. Ann. {\bf 286} (1990) 255.}\ and is given by
the coefficient $d_N$ of $q^N$ in
$$F={\prod_n(1+q^n)^{b_{odd}}\over \prod_n(1-q^n)^{b_{even}}}$$
where $b_{odd}=b_{even}=4(MP+2)$ denote the odd and even betti numbers
of $T^2\times \Sigma$.  $F$ has modular properties which allows
one to estimate the growth of the coefficient of $q^N$, following
Hardy-Ramanujan, to be
$$d_N\sim {\rm exp}(2\pi \sqrt{N(MP+2)})$$
Thus we obtain a prediction for the entropy to be
$$S=2\pi \sqrt{N(MP+2)}$$

The computation of the area of this 5 dimensional black hole by solving the
Einstein's equations in this
case gives
$$S_{BH}={A(H)\over 4}=2\pi \sqrt{NMP}$$
which agrees with what we have found in the range of validity
of the parameters $N>>M,P>>1$.

\newsec{A List of Questions}
I will list a number of questions which I believe would be interesting
to understand further.

{\bf 1}-  I have discussed some aspects of mirror symmetry.
The physical and mathematical
properties of mirror symmetry without including the D-branes
is more or less understood.  The case involving the D-branes,
which is mirror symmetry for (stable) sheaves on Calabi-Yau and is
transformed to (minimal) Lagrangian mid-dimensional cycles on the mirror
is stated in this
note. However the prediction this entails has not been checked yet.
In particular both sides of the mirror transform in this case, regardless
of the relationship between the two, deserve
further study.  Even though some aspects of stable bundles on
Calabi-Yau are known, it is rather far from a complete understanding.
The properties of minimal Lagrangian cycles and enumerative questions
in that context are even less understood.  Thus the existence of mirror
symmetry in this case may lead to many valuable mathematical insights
into both questions.

2-We have mentioned that $A-D-E$ singularities of $K3$ lead
to the appearance of the corresponding gauge group in physics.
We have also noted that some other singularities, such as a contraction
of ${\bf CP}^2$ in a Calabi-Yau threefold leads to novel physics,
not described by a conventional gauge theory.  It is thus a pretty
exciting link to develop further.  To what extent can
one classify singularity types of Calabi-Yau (and other
K\"ahler) manifolds, for three and fourfolds?  How about transitions
among manifolds mediated through singularity types?  What is a
general way to think about all manifolds at once, having in mind
their connectivity by passing through singular ones?
Among all singularities
is the appearance of $A-D-E$
singularity a rare phenomenon?  If so, what explains
the fact that we seem to live in a world with gauge symmetries?

{\bf 3}-Another issue we discussed was the counting of minimal
submanifolds.  This has some applications in the context of counting
black hole states.  There are many puzzles still to resolve in this context.
In the context of minimal 2 dimensional submanifolds
mirror symmetry gives us a way to count them in many cases of interest.
However even here there are some puzzles:  We consider a fixed
class $Q\in H_2(M,{\bf Z})$ in a Calabi-Yau threefold $M$ and ask how many
black holes exist in that class.  The predicted answer from solving the
Einstein equations is given as follows.
 Consider an arbitrary K\"ahler metric $k$
with volume 1 on Calabi-Yau $M$.  Find the K\"ahler metric which minimizes
the area of $Q$
$$V=k[{ Q}]$$
Call the minimum value $V_{min}$, and assume this is achieved for a
non-degenerate K\"ahler metric.  Then the prediction for the entropy
of the black hole \ref\pred{S. Ferrara and R. Kallosh, Phys.
Rev. {\bf D54} (1996) 1514; Phys.
Rev. {\bf D54} (1996) 1525.}, and thus the growth of moduli
of holomorphic curves in the class $Q$ is that it goes as
$$S={\rm exp}(c V_{min}^{3/2})$$
where c is a universal constant independent of Calabi-Yau.  Note
that the exponent picks up a factor of $\lambda^{3/2}$ once
we rescale $Q\rightarrow \lambda Q$.  Mirror symmetry allows
us to compute the Euler class (of an appropriate bundle) on
the moduli space of curves and that has typical growth which
upon the same rescaling of $Q$ picks up only a $\lambda$ in the exponent.
The discrepancy of this growth with that obtained in mirror computation
is presumably
because the number that mirror symmetry computes is an Euler
class, whereas the number the black hole degeneracy
predicts is the growth of cohomologies of the moduli space.  It also
suggests there must be an enormous cancellation among even and odd
cohomology states for such a dramatic change in the growth of states.  It
would be interesting to verify this.

For other types of black holes other counting problems arise.
For example, for type IIB strings  with target space being a Calabi-Yau
threefold times $R^4$ we need to count the growth in the cohomology
of the moduli space of minimal Lagrangian 3-submanifolds in a given
class $Q\in H_3(M)$.  The prediction from the black hole
side is that if we denote by $\Omega $ the holomorphic 3-form
on the Calabi-Yau and minimize
\eqn\mina{V={|\Omega(Q)|\over \sqrt{\int_M \Omega \wedge {\overline \Omega}}}}
over the moduli space of complex structure of the Calabi-Yau, assuming
that the minimum exists and does not correspond to a degenerate Calabi-Yau,
then the growth in the cohomology of moduli space of the minimal
submanifold in that class (together with a flat connection) is given by
$$S={\rm exp}(c' V_{min}^2)$$
where $c'$ is a universal constant.  In order to verify such
predictions we need to be able to count minimal Lagrangian
submanifolds.  The basic question is how to enumerate them and check
this prediction? What is the analog
of ``mirror symmetry'' which allows counting $p$ branes with $p>2$?
In fact I would conjecture, based on a few examples (not predicted
from physics) that for a Calabi-Yau of complex dimension d, if we
consider real minimal Lagrangian submanifolds of dimension d and minimize
$V$ again as given by \mina\ then the growth of the cohomology of their
moduli space (together with a flat connection) is given by
$$S={\rm exp}(c(d) V_{min}^{d-1})$$
where $c(d)$ is a universal constant depending only on $d$.
This formula is true for $d=2,1$ (in the $d=1$ case it
is vacuous and in the $d=2$ case it can be verified) and is predicted to be
true as discussed
above for $d=3$, and I am conjecturing it to be true for all $d$.
Is this true?  (Note that by mirror symmetry, this conjecture
gets transformed to counting the growth of the cohomology of moduli
of stable bundles on the mirror Calabi-Yau.)

4-We have seen many instances of dualities in physical systems and
we have explained here some of its mathematical implications.  We do
not have a deep understanding of {\it why} these dualities even exist.
Does studying the mathematical consequences of it shed any light
on this question?  In other words, why should seemingly
difficult mathematical questions find answers in terms of very
simple dual mathematical problems?  What is the mathematical meaning
of duality?

Given all this relation between physics and mathematics
one recalls Wigner's thoughts on this relationship
and in particular the ``unreasonable effectiveness of mathematics'' in solving
physical problems.  With recent developments in physics and its mathematical
implications
one may also reverse the arrow and wonder about
the unreasonable effectiveness of physics in solving mathematical problems.

I would like to thank the many collaborators I have worked
with over the years, who have greatly influenced my understanding
of the subject presented here.  I would also like to thank
Sheldon Katz for a careful reading of this manuscript and for his suggestions
for improvement.

\listrefs

\end